\documentstyle[preprint,aps]{revtex}
\newcommand {\ybco} {\rm YBa_2Cu_3O_{6+x}}

\newcommand {\lsco} {\rm La_{2-x}Sr_xCuO_4}
\newcommand {\bsco} {\rm Bi_2Sr_2CaCu_2O_8}

\newcommand {\Tc} {T_c}
\newcommand {\beq} {\begin{equation}}
\newcommand {\eeq} {\end{equation}}
\begin{document}
\draft
\title{
Anisotropic Localization Effect in Layered Materials
}
\author{Yuyao Zha$^1$ and Dongzi Liu$^2$}
\address{$^1$
Department of Physics and the Science and Technology Center of
Superconductivity, University of Illinois at Urbana-Champaign,
Urbana, IL 61801
}
\address{$^2$
James Franck Institute, University of Chicago, Chicago, IL 60637
}
\maketitle
\begin{abstract}
We investigate localization properties in the highly anisotropic and
intrinsically disordered layered material, which is analogous to high-$\Tc$
cuprates. By varying the anisotropy of the system which is parameterized by
the interlayer hopping $tp$, we find a crossover from two-dimensional (2D)
 to three-dimensional (3D) behavior at a
critical hopping amplitude $tp_c$, where a mobility edge starts to appear. We
show that below the mobility edge, anisotropic localization effect may exist
for a finite size system, when the $ab$-plane localization length is longer
than the system size and the $c$-axis localization length is shorter than the
system size. Nevertheless, we argue that such anisotropic localization can not
account for the ``semiconductor'' like behavior of the $c$-axis resistivity of
high $\Tc$ cuprates.
\end{abstract}




It has been controversial whether anisotropic
localization could exist in intrinsically disordered
layered systems such as the high-$T_c$ cuprates\cite{kotliar,ong}.
There are two interesting theoretical issues:  first, whether
it is possible at all for a three-dimensional (3D) system to be
localized in one direction and
extended in the other two directions;
secondly, if it is possible, whether this anisotropic localization effect
can account for the $c$-axis resistivity in high $T_c$
cuprates. In this paper, we address both issues.

According to the scaling theory of metal-insulator transition
\cite{anderson:scal},
for an infinite 2D system, no matter how weak the disorder is, there
is no real metallic behavior, the system is localized.
In the presence of inelastic scattering, however, when the length
scale of inelastic scattering
$L_{in}$ is smaller than the localization length $\xi$,
electron loses its  phase
coherence before it can finish the back scattering process, thus localization
can not be realized. This is the so called finite size effect, the system
has the same physical property as if it has a size of order of $L_{in}$.
In a physical system, the inelastic scattering length $L_{in}$ is a function
of temperature which decreases
as  temperature increases, thus effectively, one can mimic the effect of
temperature by changing the cutoff length of the system.
If the localization length is longer than the cutoff length, the system would
have metallic behavior; otherwise, it would behave as an insulator.
This is the reason why two-dimensional electron systems in semiconductor
heterostructures have very high mobility. One expects that
anisotropic transport behavior (such as in high $\Tc$ cuprates)
could arise in a finite-size (or at finite temperature) anisotropic system
(which has different  localization lengths along different directions)
even if the system is below the
mobility edge.

While a 2D system is always localized, it is known that there is a mobility
edge in 3D system, where metal insulator transition occurs\cite{anderson:scal}.
We model the highly anisotropic and disordered layered material
by an anisotropic tight binding lattice, which is parameterized
by hopping $t$ (which is set to be 1 throughout this paper)
in the $ab$-plane and $tp$ along the $c$-axis.
In the limit of $tp=0$, the system is equivalent to a set of disconnected 2D
 planes, and therefore has no mobility
edge. While at $tp/t=1$, the model describes an isotropic 3D system,
 there is a well defined mobility edge, whose
position is in principle a function of disorder.
It is interesting to investigate the localization properties in the
intermediate regime ($0<tp/t<1$).
Previous work \cite{qiming} on this model has shown that the mobility edge
immediately emerges once a finite $tp$ is introduced. However, in the
presence of strong disorders, the situation is more complicated.
One expects that the mobility edge appears
only when $tp$ exceeds a critical value $tp_c$ which is
a function of the disorder.
Equivalently, if one fixes the energy to be at the
band center $E=0$, and changes $tp$, one expects a metal-insulator
transition at $tp=tp_{c}$, since $E=0$ is where the extended states start to
form once the mobility edge appears. This effect is somewhat similar
to the effect studied by Clark $et~al$\cite{anderson}, in which they find
a critical hopping amplitude between two coupled Luttinger Chains,  which
separates coherent from incoherent tunneling between two chains.

In this paper, we systematically investigate the possible anisotropic
localization effect in layered materials.
The  conductances
for a cubic sample (along both $c$-axis and $ab$-plane)
are calculated by  using the Landauer-B\"{u}ttiker formula.
Our numerical results
show that upon increasing the strength of
interplane hopping through a critical $tp_c$, $c$-axis conductance
changes from semiconductor like to metallic, provided that the $ab$-plane is
sufficiently disordered (characterized by disorder strength $W$),
yet metallic. We attribute this phenomenon to the
highly anisotropic localization effect -- a metal-insulator transition
along $c$-axis for a finite-size sample.
We also calculate the finite-size
localization length along the c-axis using the transfer matrix technique.
By carrying out a finite-size scaling analysis, we found a 2D to 3D crossover
where a mobility edge starts to emerge
when $c$-axis hopping is above the same critical $tp_c$.
For highly anisotropic system (such as $tp<0.3t$), the critical $tp_c$
could be determined from
a universal relation $W/tp_c \sim 30-40$.
We discuss the relevance of
this anisotropic localization effect to the anisotropic transport
phenomena in high-$T_c$ cuprates.

In the following, we briefly outline our model and technique
to calculate the conductance.
We model our anisotropic three-dimensional system
in a cubic geometry with a finite width ($M$) square lattice
with nearest neighbor hopping. The disorder potential is modeled by
the on-site white-noise potential $V_{ijm}$ ($i,j$ denote the column and chain
indices, $m$ denotes the plane index) ranging from $-W/2$ to $W/2$.
The interplane hopping along $c$-axis $tp$ is different from the $ab$-plane
hopping amplitude $t$.
We calculate the conductance in both $ab$-plane and $c$-axis, for
different degree of anisotropy levels $tp/t \sim 0-0.5$.
The Hamiltonian of this system can be written as:
\begin{eqnarray}
{\cal H} &= &\sum_{ijm} V_{ijm}|ijm><ijm| \\
& &-t\sum_m\sum_{<ij;kl>}\left[
|ijm><klm|+c.c.\right] \nonumber \\
& &
-tp \sum_{ij}\sum_{m}\left[
|ijm><ijm+1|+c.c.\right],\nonumber
\end{eqnarray}
where $<ij;kl>$ indicates nearest neighbors on the $m$th plane. From
now on, we choose
$ab$-plane hopping amplitude $t$ as the energy unit ($t=1$ throughout the
paper).
The dc conductance at Fermi energy $E$
for a cubic sample is calculated using the
Landauer-B\"{u}ttiker formula\cite{landauer}:
\begin{equation}
 g=\frac{e^2}{h}{\rm Tr}T^{\dagger}(E)T(E),
\end{equation}
where $T(E)$ is the transmission matrix which is related to the retarded
Greens function $G^+(0,L)$ across the disordered sample (along direction $L$
in a $M\times M\times L$ sample).
The retarded single particle Greens function can be calculated using
the recursive Greens-function algorithm \cite{mackinnon}.

We present in Fig.1, the calculated resistance (1,000 sample ensemble average)
as function of
the inverse of the sample size (which could mimic the effect of temperature)
for both
$ab$-plane and $c$-axis,
with different strength of disorder W. All the calculations
are at the band center, where energy $E=0$. From Fig.1, we can see that for
$W=2, 4, 6,$ and 8, the in-plane resistance increases with $1/M$, displaying
a metallic behavior, even more so (the slope increases) as
the hopping amplitude $tp$ in the $c$-axis increases.
While the $c$-axis resistance may change from metallic
to insulating
 by changing $tp$. For example, for $W=2$, the $c$-axis resistance
$R$ is metallic-like for $tp \geq 0.05$, while for $tp \leq 0.02$, $R$
increases with sample
size $M$, the system appears to be localized along $c$ direction. In this
case, there seems to be a critical hopping amplitude $tp_{c} \sim 0.03-0.04$
that separates the localized and the metallic conduction along the
$c$-axis. As we go to higher degree of disorder, $tp_{c}$ increases.
This anisotropic localization effect (the system is only localized along
$c$-axis) does not violate the scaling theory of
localization\cite{anderson:scal}
 which states that there only
exists a single mobility edge for any disordered system even if it is highly
anisotropic (as shown in recent numerical calculations \cite{rojo}), {\it
i.e.} for large enough system, $ab$-plane would undergo the metal-insulator
transition at the same $tp_c$ as along the $c$-axis.
Below the
mobility edge, the localization length in the $ab$-plane is extremely large,
thus the system could have metallic behavior along $ab$-plane if the sample
size is smaller than the localization length. Nevertheless, if the system is
big enough, we expect the $ab$-plane would have insulating behavior\cite{c1}.
It is worth mentioning that above the mobility edge (when $tp>tp_c$), the
calculated conductance in both $ab$-plane and $c$-axis increase linearly with
the increasing sample size $M$, indicating typical 3D Drude conductance
behavior.

The finite size correction to the conductivities of an anisotropic system
 has been calculated by W\"{o}lfle and Bhatt,\cite{wolfle} from evaluating the
maximally crossed diagrams by assuming both anisotropic single particle
energy dispersions and anisotropic impurity scatterings. Their results
show that backscattering  gives rise to a diffusive correction to
the metallic conductivity, and that the effects of anisotropy near the weak
coupling (metallic) fixed
point can be completely absorbed into an anisotropic diffusion coefficient.
More recent work by Szott $et~al$ has
microscopically derived the anisotropic diffusion constants from the
anisotropic impurity scattering rates for the superlattice semiconductor, and
concluded that the conductivity for superlattices still has the diffusive
form.\cite{szott}
Nevertheless, these above work did not address the 2D-3D transition found in
our numerical calculations. To be more specific,
it is shown that the contribution from
backscattering (weak localization correction)
 reduces the conductivity differently in 2D and 3D,\cite{pklee}
\begin{mathletters}
\begin{eqnarray}
\sigma_{3D}(M)&=&\frac{ne^2\tau}{m^*}-\frac{e^2}{\hbar\pi^2}\left[\frac{1}{l}-\frac{1}
{M}\right]
\\
\sigma_{2D}(M)&=&\frac{ne^2\tau}{m^*}-\frac{e^2}{\hbar\pi^2}{\rm
ln}\left[\frac{M}{l}\right]
\end{eqnarray}
\end{mathletters}
where $l$ is the mean free path. Therefore, we expect that in our anisotropic
layered
system, the weak localization correction would have a crossover from
logarithmic
behavior to $1/M$ as $tp$ increases from 0 (pure 2D system)  to 1 (isotropic
3D system).
On the other hand, our numerical data show that
above the mobility edge (when $tp>tp_c$), the
calculated conductance in both $ab$-plane and $c$-axis increase linearly with
the increasing sample size $M$, exhibiting a typical 3D Drude conductance.
For large sample sizes, the breaking away from linear behavior in
our data compare favorably to a negative weak localization correction
in the conductivity. However, strong finite size effect and statistical
fluctuations do not allow us to make a quantitative comparison with the
analytic results of Refs \cite{wolfle} and \cite{szott}.

In order to study the transition between 2D and 3D behaviors
 by tuning $tp$ in our anisotropic tight-binding model, we need to calculate
the localization length directly.
We now introduce the technique to calculate the localization length
in a long square bar geometry ($M\times M\times L$) along  the $c$-axis.
For a specific energy $E$, a transfer matrix $T_{m}$ can be easily set up
mapping the wavefunction amplitudes at plane ${m-1}$ and $m$ to those
at plane $m+1$, {\it i.e.}
\begin{equation}
\left( \begin{array}{c} \psi_{m+1} \\ \psi_{m} \end{array} \right) =
T_{m}
\left( \begin{array}{c} \psi_{m} \\ \psi_{m-1} \end{array} \right) =
\left( \begin{array}{cc} \frac{H_{m}-E}{tp} & -I \\ I & 0 \end{array} \right)
\left( \begin{array}{c} \psi_{m} \\ \psi_{m-1} \end{array} \right) ,
\end{equation}
where $H_m$ is the Hamiltonian for the $m$th plane, $I$ is a $M^2\times
M^2$ unit matrix.
Using a standard iteration algorithm \cite{ldz:bloc}, we can
calculate the Lyapunov exponents for the transfer matrix $T_{m}$.
The localization length $\lambda_M(E)$ for energy
$E$ at finite width $M$ is then given by the inverse of the smallest
Lyapunov exponent. We choose sample length $L (\approx 1000)$
to be long enough so that the
self-averaging effect automatically takes care of the ensemble statistical
fluctuations. In our calculation, we fix the energy at $E=0$ and change the
hopping amplitude $tp$ for sample size $M=4\sim 16$.

To obtain the thermodynamic properties of the system, we carry out a
finite-size scaling analysis of the renormalized localization length
$\lambda_M/M$, which satisfies a simple scaling law. As shown in
Fig.2,  all $\lambda_M/M$ data for different disorder and
hopping amplitude $tp$ fall on the same curve which can be written as:
\begin{equation}
\lambda_M(W,tp)/M=f[\xi(W,tp)/M]
\end{equation}
where $\xi(W,tp)$ is a characteristic length which depends on $W$ (or energy
$E$) but not on $M$. Similar to the scaling analysis for an isotropic 3D
system \cite{kramer}, the scaling function has two branches. The scaling
parameter associated with the lower branch corresponds to the thermodynamic
localization length, {\it i.e.} $\xi = \lambda_{\infty}$. The one associated
with the upper branch belongs to the extended regime and identifies with the
resistivity of the system \cite{kramer}. The scaling parameter should diverge
at the critical point which separates the two branches in the scaling function
indicating a metal-insulator transition.
The inset of Fig.2 plot the scaling parameter $\xi$ as a
function of $W/tp$ for all
the $W$ and $tp$ shown in the scaling curve. We can see that for the highly
anisotropic system we considered ($tp/t<0.5$), the scaling
parameter is a universal function of $W/tp$.
At $W/tp \sim 30-40$, $\xi$ diverges,
showing a typical Anderson metal-insulator transition. We now compare to
Fig.1, where the critical hopping amplitude separating the metallic and
localized regimes for $W=2,4,6,8$ all seem to be pinned at $W/tp_c \sim
30-40$. Thus we
argue for this highly anisotropic
 tight-binding model ($tp<0.5$), a typical metal-insulator
transition  occurs at certain $tp_{c}(W)$, where $W/tp_{c}$ is a universal
number independent of hopping $tp$ and on-site disorder $W$.

Here we should emphasize that the relation $W/tp_c\sim 30-40$ is only true
for highly anisotropic system where $tp<<t$. (We fix $t=1$ in this paper.)
However, for systems with modest anisotropy, namely $0.5< tp/t <1$, we do not
expect that the relation $W/tp_c\sim 30-40$ to be relevant.
In fact, we find that $W/tp_c\approx 16$ for isotropic
system ({\it i.e.} $tp=t$), which is consistent with previous
works\cite{qiming,kramer}. For the strong anisotropic system we have studied,
where $tp/t <0.5$, the system is closer to 2D, the mobility edge
is mostly determined by the small amount of interlayer hopping $tp$,
while
the in-plane hopping $t$ plays a much weaker role in determining the
critical point of $W/tp_c$ other than setting the energy scale of both $W$
and $tp$. As the system becomes less anisotropic, $tp$ and $t$ become
compatible, therefore should both determine
the mobility edge.
Thus we find
for $tp/t <0.5$, the mobility edge appears at $W/tp_c \sim30-40$
and $W/t \sim 2-8 $; when anisotropy decreases, at $0.5<tp/t<1$,
it is only natural that these two ratios merge closer, and eventually,
at $tp=t$, we have $W/tp_c=W/t \sim 16$.
Another way to understand this is to write the critical ratio $W/tp_c$ as
function of $tp/t$ which characterizes the anisotropy of the system,
 we find $W/tp_c= 30-40$ at $tp/t \ll 1$, and
$W/tp_c=16$ at $tp/t=1$. We expect a smooth crossover in the intermediate
anisotropy regime of $0.5<tp/t<1$. Our results show that $W/tp_c$ is overall
$t$-dependent, while in the strong anisotropic limit of $tp/t<0.5$, the
$t$-dependence of $W/tp_c$ becomes much weaker.

In conclusion, we observe a crossover from 2D to 3D behavior in the highly
anisotropic layered system when the interlayer hopping amplitude exceeds a
critical value.
This crossover behavior of localization from 2D to 3D has been studied
by various researchers, such as in Ref. 4. However, we have investigated the
non-trivial consequences of this crossover behavior by directly calculating
the conductance of finite-size samples. Anisotropic transport properties (such
as temperature dependence) could arise from the anisotropic localization
properties.
We have shown that even when the system is below the mobility
edge, in the presence of inelastic scattering which
provides a cutoff length scale, it is possible to observe localization in the
$c$-direction (if the localization length along $c$-axis is shorter than the
cut-off length)
and metallic behavior in the $ab$-plane (if the localization length in the
$ab$-pane is longer than the cut-off length) for highly anisotropic
systems.

A natural applicable system where the above anisotropic localization may occur
is the layered high $T_c$ cuprates, where the $c$-axis resistivity
indeed shows a ``semiconductor-like'' upturn at low temperatures for $\bsco$
and underdoped $\lsco$ and $\ybco$, while the upturns disappear upon doping.
Nevertheless,
optical far-infrared and Raman experiment\cite{cooper2} suggest that the
$c$-axis scattering rate $1/\tau_c$ for typical $\ybco$
is of the order of 100meV, while
the $c$-axis hopping amplitude $tp \sim 10meV$. Thus it appears that
the ratio of disorder (should not be larger than $1/\tau_c$ if there is any)
and $c$-axis hopping $W/tp < W/tp_c$.
 Moreover, the typical $ab$-plane hopping energy
is of the order of 100meV, leaving $W/t < 1$. Therefore, these cuprates are
 well above the mobility
edge, and the $c$-axis transport should be on the metallic side.
Details of the $c$-axis scattering rate and the hopping rate for
each individual cuprate at different dopings are listed in Ref.\cite{zha}.
Furthermore, unlike the simple model we present here, there is a
strong $ab$-plane fluctuation which increases with temperature, unless at
very low $T$, this planar fluctuation will destroy the phase coherence so
that back scattering can not occur\cite{ong}.
Therefore, we argue that the $c$-axis
resistivity of high-$T_c$ cuprates can not be associated with the anisotropic
localization phenomena we discussed above. An reasonable understanding of
the $c$-axis resistivity data is given in the recent work of Ref.\cite{zha}.
Nevertheless, this does not exclude the possibility that localization (either
$c$-axis or $ab$-plane) may occur in some extremely dirty samples or films or
at doping concentration in closely proximity of the insulating phase of
cuprates.

We acknowledge stimulating conversations with A.J. Leggett, P. Philips,
K. Levin, J. Miller, and
Q.M. Li. This work is supported by NSF-DMR-91-20000 and 94-16926 through the
Science and Technology Center of Superconductivity.


\begin{figure}
\caption{
Sample size dependent resistance for a cubic sample for different disorder
strength (a) $W=2$; (b) $W=4$; (c) $W=6$; (d) $W=8$.
\label{c}}
\end{figure}

\begin{figure}
\caption{The scaling function, $\lambda_M/M$ vs $\xi/M$.
The $\xi(W,tp)$ shown in the
inset turns out to
be a universal function of $W/tp$.}
\end{figure}


\begin{references}
\bibitem{c1}
Strictly speaking, one can only discuss metal-insulator transition in the
thermodynamic limit where insulating states have finite localization length.
 Practically, however, one could investigate the localization properties using
the finite size
technique. By probing the sample size dependence of the transport properties
for large enough sample, one could distinguish localized states from extended
ones
according to the arguments given in Ref.3.  In doing so, however, it
is important to make sure that the sample size is big enough so that the
asymptotic behavior is reached, because if the sample size is much smaller
than the
thermodynamic localization length, even a localized state
would have a metallic behavior.
\end{references}
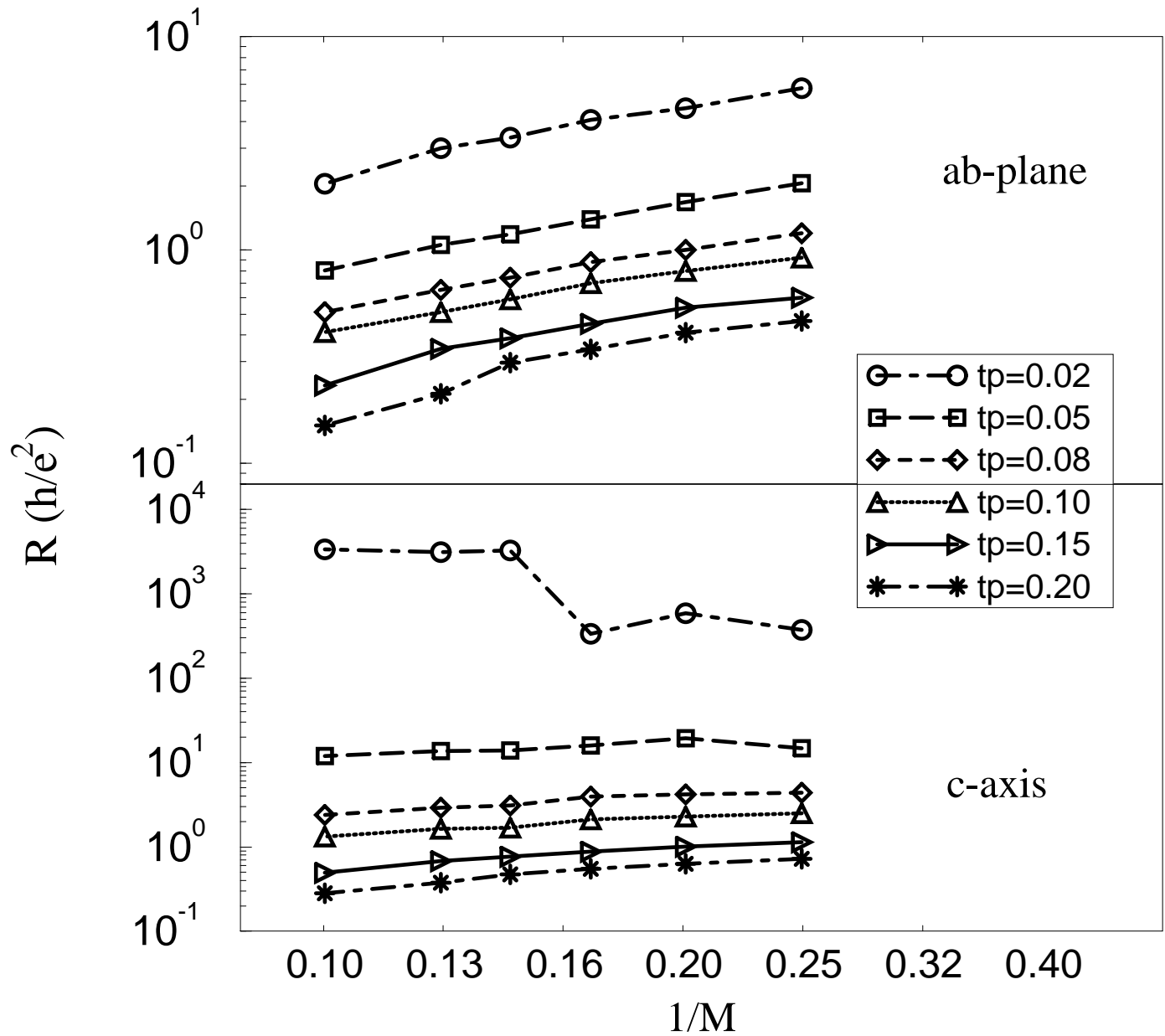
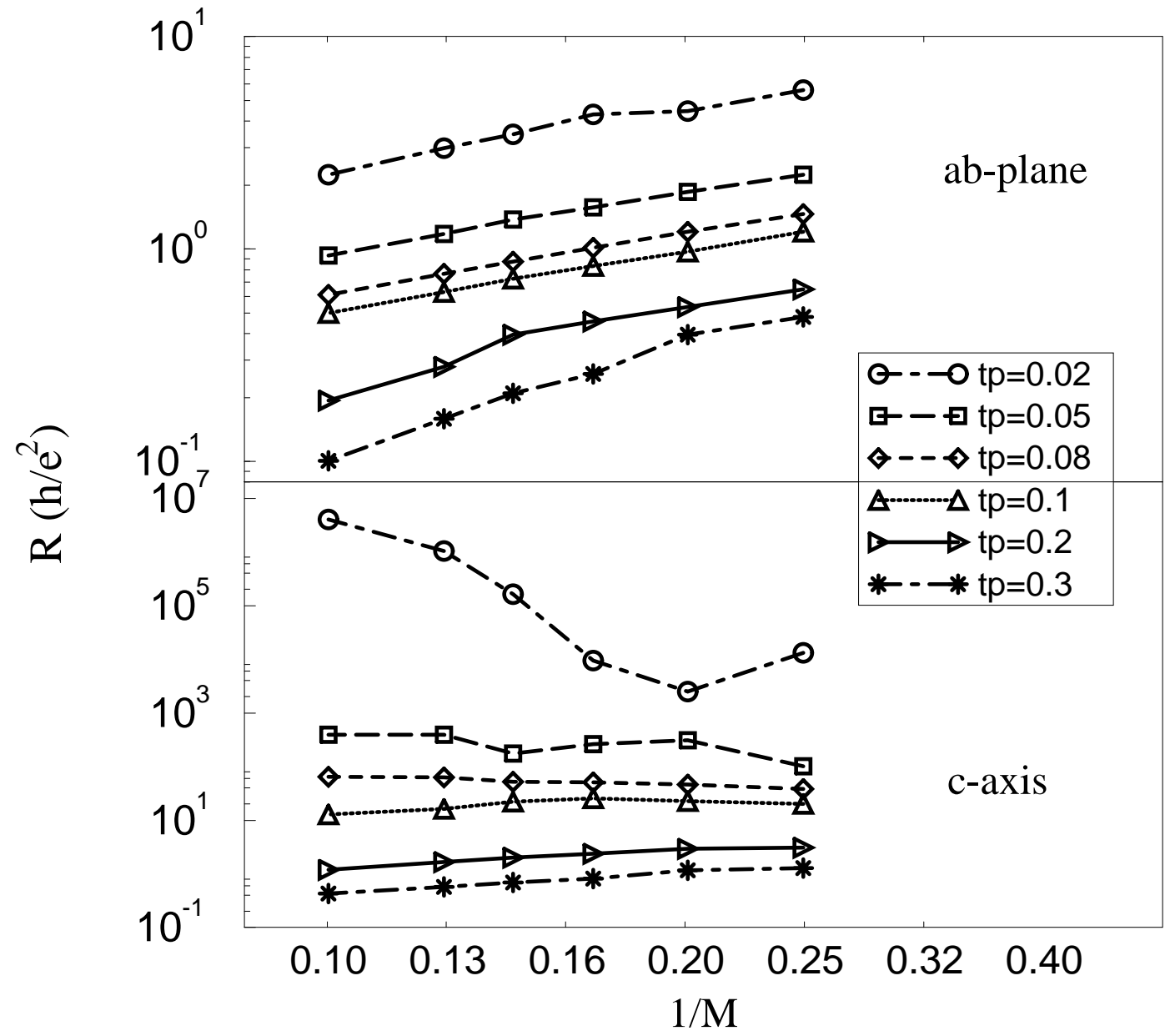
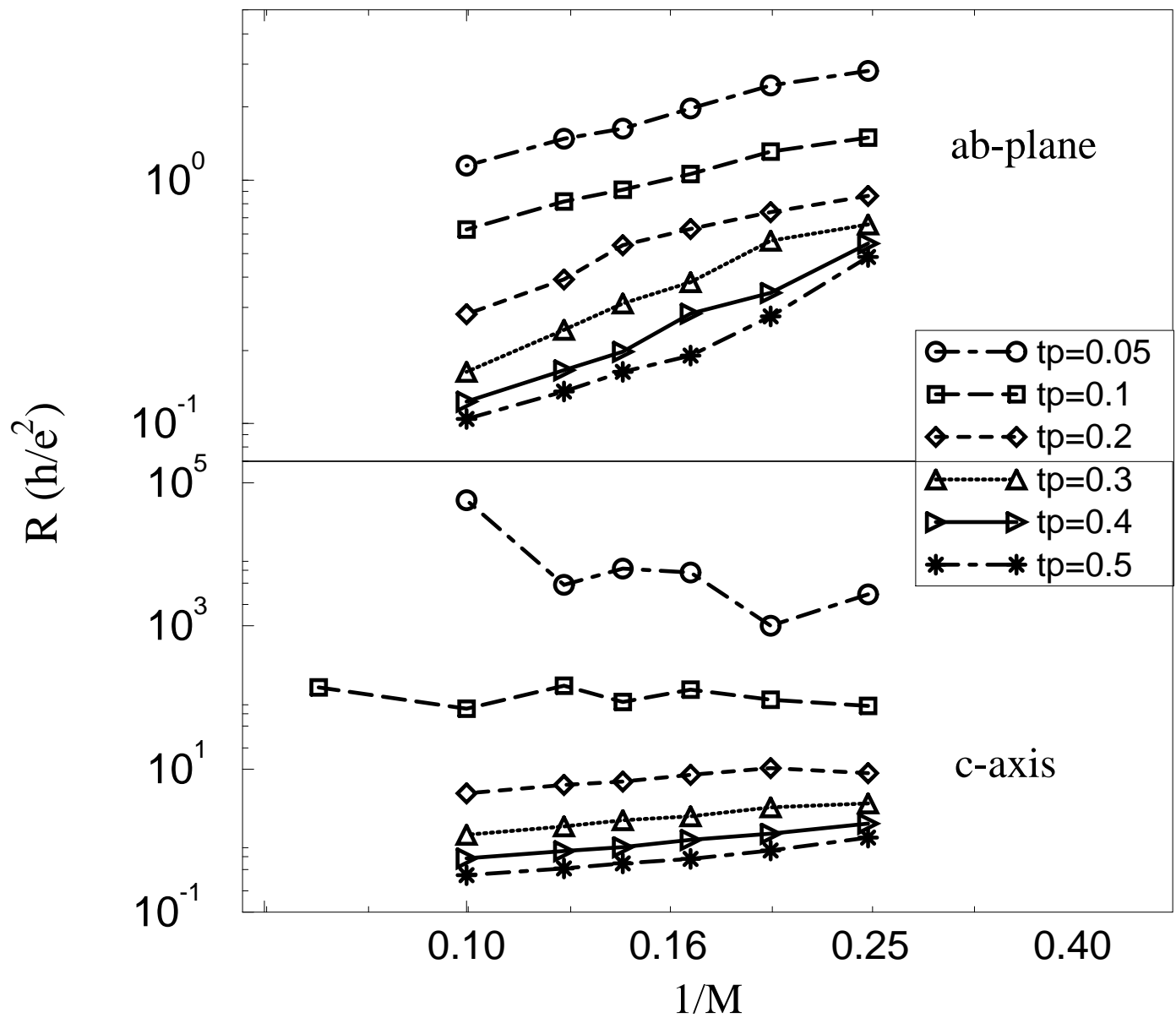
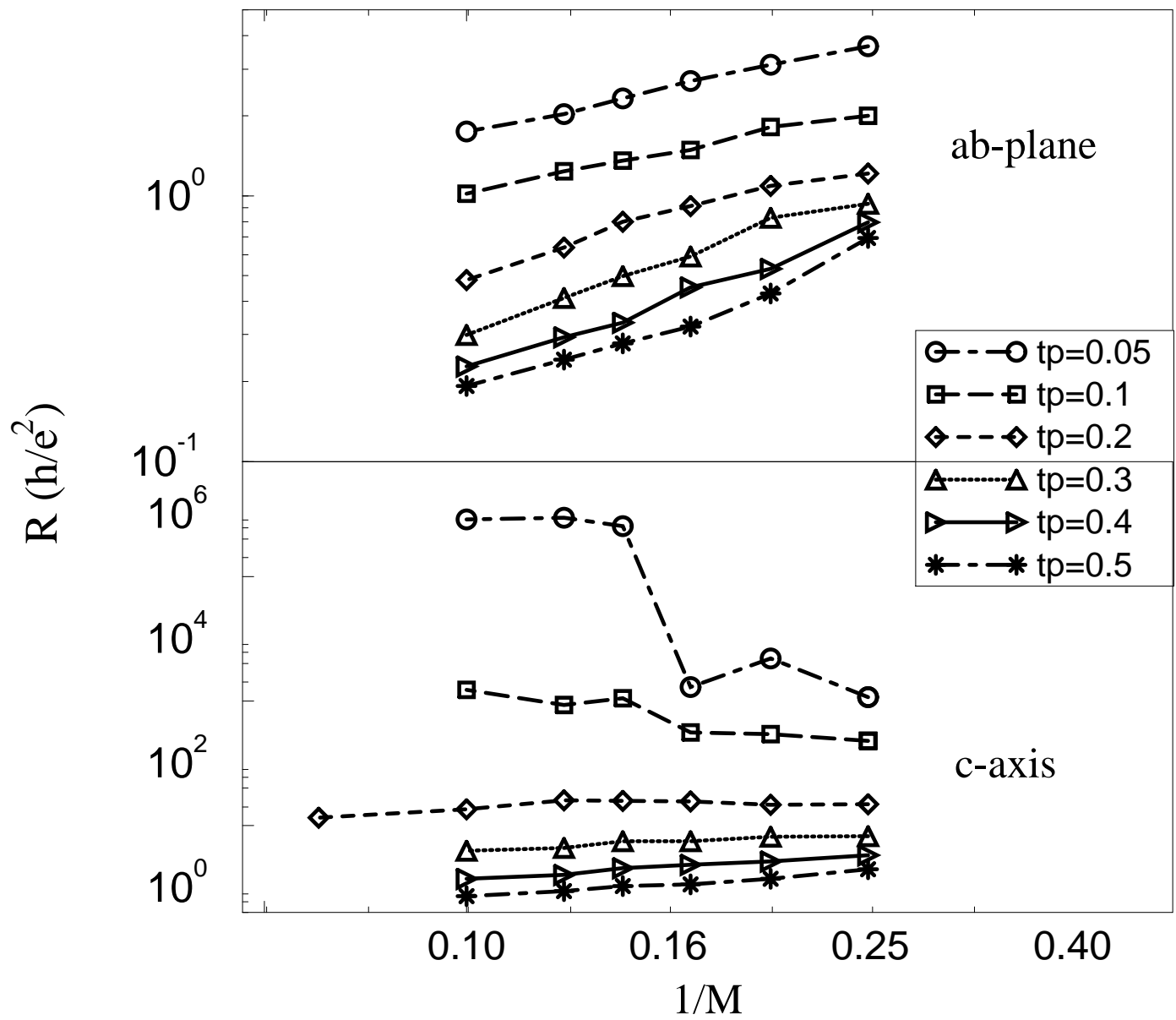
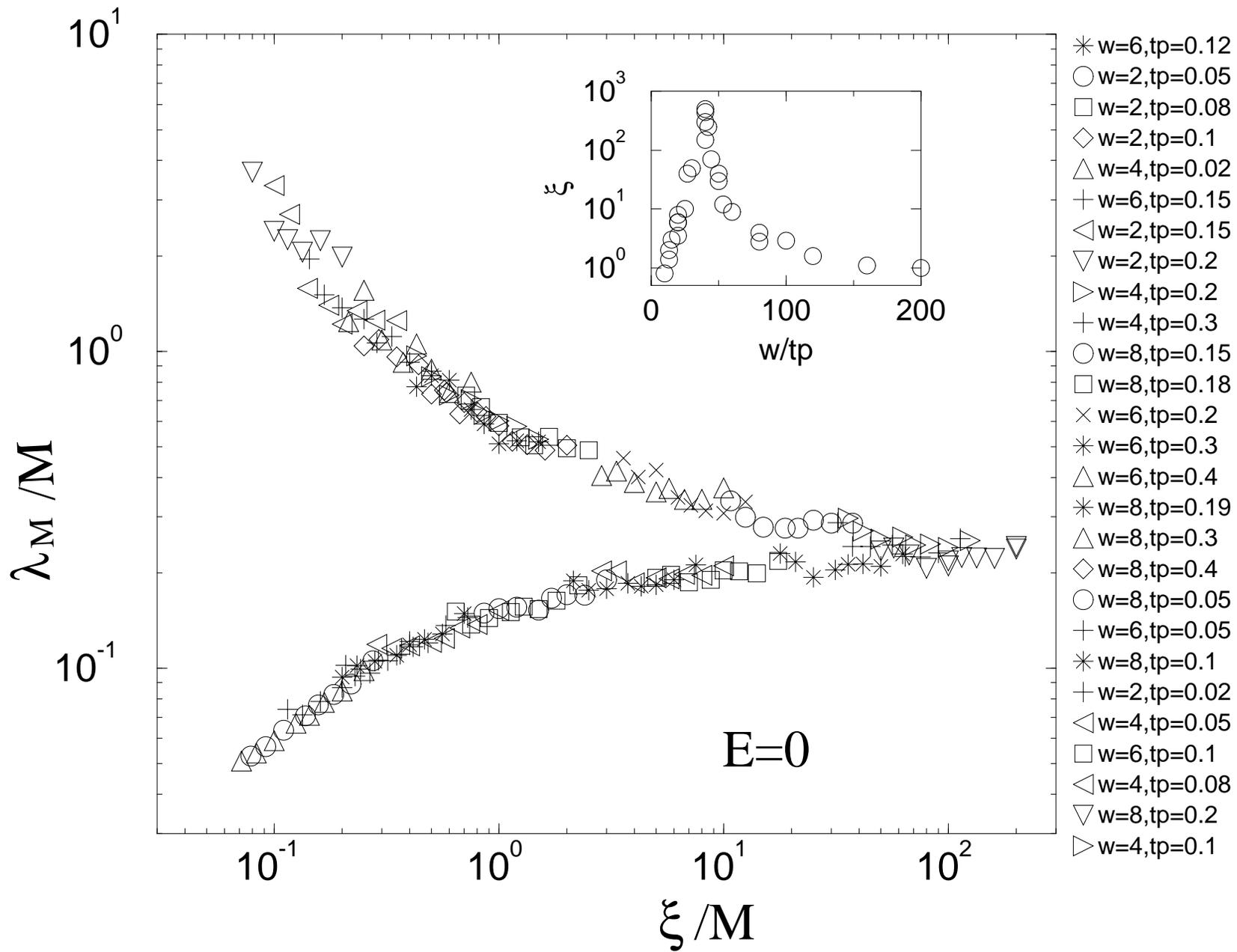
\end{document}